\input{aipcheck}

\documentclass[
    ,final            
  ]
  {aipproc}

\layoutstyle{6x9}

\begin{document}

\title{Five Dimensional Minimal Supergravities \\ and Four Dimensional Complex Geometries}


\author{Jai Grover}{
  address={DAMTP, CMS, University of Cambridge, Wilberforce Road, Cambridge, CB3 0WA, UK}
}

\author{Jan B. Gutowski}{
  address={Department of Mathematics, King's College London, Strand, London WC2R 2LS, UK}
}

\author{Carlos A.R. Herdeiro}{
  address={DF-FCUP e CFP, Universidade do Porto, Rua do Campo Alegre, 687, 4169-007 Porto, Portugal}
}

\author{Wafic Sabra}{
  address={Centre for Advanced Mathematical Sciences and Physics Department, American University of Beirut, Lebanon}
}

\begin{abstract}
We discuss the relation between solutions admitting Killing spinors of minimal supergravities in five dimensions and four dimensional complex geometries. In the ungauged case (vanishing cosmological constant $\Lambda=0$) the solutions are determined in terms of a hyper-K\"ahler base space; in the gauged case ($\Lambda<0$) the complex geometry is K\"ahler; in the de Sitter case ($\Lambda>0$) the complex geometry is hyper-K\"ahler with torsion (HKT). In the latter case some details of the derivation are given. The method for constructing explicit solutions is discussed in each case.\end{abstract}

\maketitle


\section{Introduction}

In the last few years it has been realised that there are some beautiful connections between solutions of minimal supergravities in five dimensions admitting Killing spinors and complex geometries. Moreover, these connections have been useful in finding qualitatively new black hole solutions. The first case to be considered was the minimal ungauged five dimensional supergravity, i.e with no cosmological constant ($\Lambda=0$). It was shown by Gauntlett et al \cite{Gauntlett:03} that the most general stationary solution admitting Killing spinors is defined by a four dimensional \textit{hyper-K\"ahler} base space and a set of constraint equations. Using this construction, Elvang et al \cite{Elvang:04} obtained the supersymmetric black ring. The second case to be considered was the minimal gauged five dimensional supergravity, i.e with negative cosmological constant ($\Lambda<0$). It was shown by Gauntlett and Gutowski \cite{Gauntlett:03b} that the most general stationary solution admitting Killing spinors is defined by a four dimensional \textit{K\"ahler} base space and a set of constraint equations. Using this construction, Gutowski and Reall found the supersymmetric $AdS_5$ black holes \cite{Gutowski:04}. It is then natural to ask: i) Is there any interesting relation with complex geometry in five dimensional minimal \textit{de Sitter} supergravity (i.e with positive cosmological constant $\Lambda>0$)? ii) Can we use the resulting structure to find interesting solutions? In the following we will discuss how indeed the answer to i) is yes and give some hints towards answering ii).

\section{Minimal Ungauged Supergravity in $D=4$}

Supersymmetric black holes are interesting gravitational objects. They are classically and semi-classically stable; in many cases there is a no-force condition which allows for a multi-object configuration. An early example is obtained in minimal $\mathcal{N}=2$, $D=4$ supergravity, whose bosonic sector has action
\[ \mathcal{S}=\frac{1}{16\pi G_4}\int d^4 x\sqrt{-g}\left(R-\frac{F^2}{4}\right) \ ; \]
the ansatz
\[ds^2=-H(x)^{-2}dt^2+H(x)^2ds^2_{{E}^3} \ , \ \ \ \ \ \ A=H(x)^{-1}dt \ , \]
reduces the full non-linear Einstein-Maxwell system to a single harmonic equation on the Euclidean 3-space $E^3$
\[ \Delta_{{E}^3} H(x)=0 \ . \]
This is the well known Majumdar-Papapetrou solution, and taking $H(x)$ to be a multi-centred harmonic function yields a multi black hole solution. Physically, the exact linearisation of the supergravity equations, yielding a superposition principle, is associated to the exact balance of electrostatic repulsion and gravitational attraction between any pair of black holes. But it is also associated to supersymmetry. Indeed, Tod \cite{Tod:83} showed that the most general stationary solution of this theory admitting Killing spinors, i.e. a non-trivial solution of 
\[ D\epsilon-\frac{1}{4}F_{ab}\Gamma^{ab}\Gamma\epsilon=0  \ , \]
falls into the class of Israel-Wilson-P\'erjes metrics (which have the modulus of charge equal to mass). Since the most general black hole solution in this class is the Majumdar-Papapetrou solution one concludes the latter is the most general static (indeed the most general stationary) supersymmetric black hole solution of $\mathcal{N}=2$, $D=4$ supergravity.

\section{Minimal Ungauged Supergravity in $D=5$}

It follows from the above that Einstein-Maxwell theory (seen as the bosonic sector of supergravity) does not admit any supersymmetric \textit{rotating}, asymptotically flat, black holes. And for some time it was even doubtful that such an object could exist; indeed supersymmetry is a statement of stability and rotation is normally associated to the instabilities which arise from the existence of an ergo-region (Penrose process and superradience). However one such solution was found in five dimensional minimal ungauged supergravity, whose bosonic sector has action
\[ \mathcal{S}=\frac{1}{16\pi G_5}\int d^5x\left[\sqrt{-g} \left(R-F^2\right)-\frac{2}{3\sqrt{3}}A\wedge F\wedge F\right] \ ; \]
the ansatz
\[ds^2=-H(x)^{-2}(dt+\omega)^2+H(x)ds^2_{{E}^4} \ , \ \ \ \ \ \ A=\frac{\sqrt{3}}{2}H(x)^{-1}\left(dt+\omega\right) \ \ , \]
reduces the supergravity equations to the constraints \cite{Gauntlett:98}
\[ \Delta_{{E}^4} H(x)=0 \ , \ \ \ \ d\omega=-\star^{(4)}d\omega \ , \]
where $\star^{(4)}$ is the Hodge dual on the Euclidean 4-space $E^4$. This is known as the BMPV solution \cite{BMPV:97}. Taking $H(x)$ to be a multi-centred harmonic function, and an appropriate choice for $\omega$, yields a solution with multiple black holes in an asymptotically flat spacetime. Physically, the exact linearisation of the supergravity equations is again associated to the exact balance between electromagnetic and gravitational forces between any pair of black holes. But now, besides electric we will have magnetic effects (spin-spin and magnetic dipole-dipole forces). And again it is also associated to supersymmetry. Indeed, Gauntlett et al \cite{Gauntlett:03} showed that the most general stationary solution of this theory admitting Killing spinors, i.e. a non-trivial solution of 
\[ \left[D_{\alpha}+\frac{1}{4\sqrt{3}}\left(\Gamma^{\ \beta \gamma}_{\alpha}-4\delta^{\beta}_{\alpha}\Gamma^{\gamma}\right)F_{\beta \gamma}\right]\epsilon^a=0 \ , \]
is of the form
\[ds^2=-f^{2}(dt+\omega)^2+f^{-1}ds^2_{\mathcal{M}} \ , \ \ \ \ F=\frac{\sqrt{3}}{2}d(f[dt+\omega])-\frac{G^+}{\sqrt{3}} \ , \]
and that the method to construct explicit examples is the following:
\begin{itemize}
\item[1)] Choose $\mathcal{M}$ to be a 4 dimensional hyper-K\"ahler manifold;
\item[2)] Decompose $f d\omega=G^++G^-$;
\item[3)] Solve
\[dG^+=0 \ , \ \ \ \ \ \Delta f^{-1}=\frac{2}{9}(G^+)^2 \ . \]
\end{itemize}
Taking $G^+=0$ and $\mathcal{M}=E^4$, we find the BMPV (multi) black hole solution. But this choice also includes other, qualitatively different solutions, namely maximally supersymmetric G\"odel type universes and black holes in G\"odel type universes \cite{Gauntlett:03,Herdeiro:03}.

Taking $G^+\neq 0$ one can find supersymmetric black rings. But note that in this case one does not find a harmonic equation any longer, but rather a Poisson type equation, to which $G^+$ is the source. Thus, the superposition principle is, in general, lost. And in fact there is no solution with multiple black rings where these can be placed at arbitrary positions, as for the multiple black holes seen above. Nevertheless, writing $\mathcal{M}$ as a Gibbons-Hawking space one can construct multiple concentric black rings \cite{Gutowski:04b}.

\section{Minimal Gauged Supergravity in $D=5$}

A negative cosmological constant is introduced by moving to minimal gauged supergravity in five dimensions. This theory is particularly relevant because it is related by the AdS/CFT duality to the well understood $\mathcal{N}=4$, $D=4$ Super Yang Mills theory. Hence one might expect to give a microscopic interpretation to the black hole solutions in this theory using the duality. A generic analysis of the Killing spinor equation
\[ \left[D_{\alpha}+\frac{1}{4\sqrt{3}}\left(\Gamma_{\alpha}^{\ \beta \gamma}-4\delta_{\alpha}^{\beta}\Gamma^{\gamma}\right)F_{\beta \gamma}\right]\epsilon^a-g\epsilon^{ab}\left(\frac{\Gamma_{\alpha}}{2}-\sqrt{3}A_{\alpha}\right)\epsilon^b=0 \ , \]
shows that all susy solutions with a timelike Killing vector field are of the form \cite{Gauntlett:03b}:
\[ds^2=-f^{2}(dt+\omega)^2+f^{-1}ds^2_{\mathcal{M}} \ , \ \ \ \ F=\frac{\sqrt{3}}{2}d(f[dt+\omega])-\frac{G^+}{\sqrt{3}}+\sqrt{3}gf^{-1}J  \ , \]
and that the method to construct explicit solutions is the following:
\begin{itemize}
\item[1)] Choose $\mathcal{M}$ to be a 4 dimensional K\"ahler manifold;
\item[2)] Compute
\[f=-\frac{24g^2}{R} \ , \ \ \ \ \ \  \ G^+=-\frac{1}{2g}\left[\mathcal{R}+\frac{R}{4}J\right] \ , \]
which determines completely $f$ and $G^+$ in terms of the properties of the K\"ahler space: its Ricci scalar, $R$, its Ricci form, $\mathcal{R}$, and its K\"ahler form, $J$; solve also
\[ \Delta f^{-1}=\frac{2}{9}(G^+)^{mn}(G^+)_{mn}-gf^{-1}(G^-)^{mn}J_{mn}-8g^2f^{-2} \ ,  \]
which determines the components of $G^-$ that have a non trivial contraction with the K\"ahler form;
\item[3)] Solve the constraint
\[f d\omega=G^++G^- \ , \]
which determines the remaining components of $G^-$.
\end{itemize}
Taking $\mathcal{M}$ to be Bergamann space, which can be written as \cite{Figueras:06}
\[ ds^2_\mathcal{M}=d\sigma^2+\frac{\sinh^2g\sigma}{4g^2}\left(\frac{dx^2}{H(x)}+H(x)d\psi^2+\cosh^2g\sigma(d\phi+xd\psi)^2\right) \ , \]
with $H(x)=1-x^2$, one finds $AdS_5$. The Gutowski-Reall black hole \cite{Gutowski:04} and the Chong et al. black hole \cite{Chong} are found by taking more general quadratic and cubic polynomials \cite{Figueras:06}. Note that for arbitrary $H(x)$ the above metric is K\"ahler; but the supersymmetry constraints impose $(H^2H'''')''=0$, where primes denote $x$ derivatives, which shows that a given base space might not give rise to a five dimensional solution. Also, a given base space might give rise to a family of solutions with an infinite number of parameters \cite{Figueras:06}.

It is worth noting that the supersymmetric $AdS_5$ black holes found using this construction \textit{must} rotate, which is similar to what happens in three \cite{Coussaert:93} and four \cite{Kostelecky:95} dimensions.

\section{Minimal de Sitter Supergravity in $D=5$}
It is well known that de Sitter superalgebras have only non-trivial representations in a positive-definite Hilbert space in two dimensions \cite{Pilch:85,Lukierski:85}. Nevertheless one can take the perspective of \textit{fake supersymmetry}, in analogy to the recently explored Domain Wall/Cosmology correspondence \cite{Skenderis:06}: that there is a special class of solutions in a gravitational theory with a positive cosmological constant admitting ``pseudo-Killing spinors''. Thus, fake supersymmetry becomes a solution generating technique, as we shall explain. Note that, nevertheless, a relation to fundamental theory still exists via compactifications of the IIB* theory \cite{Hull:98,Sabra:03}.

We now follow closely \cite{Grover:08}. The action for minimal de Sitter supergravity in $D=5$ is 
\[ \mathcal{S}=\frac{1}{4\pi G_5}\int \left(\frac{1}{4}(^5R-\chi^2)\star
1-\frac{1}{2}F\wedge \star F-\frac{2}{3\sqrt{3}}F\wedge F\wedge A
\right) \ , \]
and the Killing spinor equation is 
\begin{eqnarray}
\label{eqn:grav}
 \bigg[\partial_M +
{1\over4}\Omega_{M,}{}^{N_1 N_2}\Gamma_{N_1 N_2} - {i \over
4\sqrt{3}} F^{N_1 N_2} \Gamma_M \Gamma_{N_1 N_2} +{3i \over
2\sqrt{3}} F_{M}{}^{N} \Gamma_{N} \nonumber \\ + 
\chi({i\over4\sqrt{3}}\Gamma_{M} - {1\over2}A_{M}) \bigg]
\epsilon =0 \ . \nonumber
\end{eqnarray}
The sense in which fake supersymmetry can be used as a solution generating technique is the following. If a non-trivial solution of the (pseudo) Killing spinor equation exists and the gauge field equations are satisfied, the integrability conditions of the former place constraints on the Ricci tensor. For the solutions we consider here, for which the Killing spinor generates a timelike vector field, these constraints are equivalent to the Einstein equations. Note this would not be so for the null case, for which the Killing spinor generates a null vector field.

Let us now give some details of the derivation of the (fake) supersymmetry constraints. The basic principle is to assume the existence of, at least, one non-trivial (pseudo) Killing spinor. This puts constraints on the spin connection and gauge field. In practice we use spinorial geometry techniques \cite{Gillard:04}. That is, we take the space of Dirac spinors to be the space of complexified forms on $E^2$, which is spanned over the space of complex numbers by $\{ 1, e_1, e_2, e_{12} \}$ where
$e_{12}=e_1 \wedge e_2$. The action of
complexified $\Gamma$-matrices on these spinors is given by

\begin{eqnarray} \Gamma_{\alpha} = \sqrt{2} e_\alpha \wedge \ , \qquad
\Gamma_{\bar{\alpha}} = \sqrt{2} i_{e^\alpha}\ , \nonumber \end{eqnarray}
for $\alpha=1,2$,
and $\Gamma_0$ satisfies
\[ \Gamma_0 1 = -i1 , \quad \Gamma_0 e^{12} = -ie^{12} , \quad
\Gamma_0 e^j = ie^j \ , \qquad \ j =1,2 \ , \]
where we work with an oscillator basis in which the spacetime metric is
\[
ds^2 = -({\bf{e}}^0)^2 +2 \delta_{\alpha \bar{\beta}} {\bf{e}}^\alpha {\bf{e}}^{\bar{\beta}} \ .
\]
The Killing spinor can be put in a canonical form using the $Spin(4,1)$ transformations. The canonical form is $\epsilon=h 1$, where $h$ is a function, if it originates a timelike vector and $\epsilon=1+e_1$ if it originates a null vector. We will be interested in the former case. 

Defining a 1-form $V={\bf e}^0$ and introducing a $t$ coordinate such that the dual vector field is $V=-\partial/\partial t$, a computation shows that the frames take the form
\[ {\bf e}^0=dt+\frac{2\sqrt{3}}{\chi}\mathcal{P}+e^{\frac{\chi t}{\sqrt{3}}}\mathcal{Q} \ , \qquad {\bf e}^\alpha=e^{-\frac{\chi}{2\sqrt{3}}t}\hat{\bf e}^{\alpha}  \ , \]
where
\[ \mathcal{L}_V \hat{\bf e}^{\alpha}=0 \ , \qquad \mathcal{L}_V \mathcal{Q}=0 \ , \qquad \mathcal{L}_V \mathcal{P}=0 \ . \]
We refer to the 4-manifold with $t$-independent metric 
\[ ds^2_\mathcal{M}=2\delta_{\alpha\bar{\beta}}\hat{\bf e}^\alpha \hat{\bf e}^{\bar{\beta}} \ , \]
as the ``base space'' $\mathcal{M}$. It follows that part of the geometrical constraints imposed by the Killing spinor equation are equivalent to $d J^i = -2 \mathcal{P} \wedge J^i$ for $i=1,2,3$ where 
\begin{eqnarray}
\label{hyperhermit2}
J^1 = {\hat{\bf{e}}}^1 \wedge {\hat{\bf{e}}}^2 +  {\hat{\bf{e}}}^{\bar{1}} \wedge  {\hat{\bf{e}}}^{\bar{2}} \ , \qquad
J^2 = i  {\hat{\bf{e}}}^1 \wedge  {\hat{\bf{e}}}^{\bar{1}} +  i  {\hat{\bf{e}}}^2 \wedge  {\hat{\bf{e}}}^{\bar{2}} \ ,
\qquad
J^3 = -i {\hat{\bf{e}}}^1 \wedge {\hat{\bf{e}}}^2 +i  {\hat{\bf{e}}}^{\bar{1}} \wedge  {\hat{\bf{e}}}^{\bar{2}} \ , \nonumber
\end{eqnarray}
defines a triplet of anti-self-dual almost complex structures on $\mathcal{M}$ which satisfy the algebra of the imaginary unit
quaternions. Thus, $\mathcal{M}$ is hyper-K\"ahler with torsion, HKT; in other words, the almost complex structures are preserved by a connection with torsion:
\[\nabla^+J^i=0 \ , \qquad \Gamma^{(+)}{}^i{}_{jk}=\{{}^i_{jk}\} + H^i{}_{jk}\ ,\]
where the torsion  is $H=\star_4\mathcal{P}$, and $\star_4$ is the Hodge dual on $\mathcal{M}$. 

The bottom line is that the most general solution of five dimensional minimal de Sitter supergravity admitting a (pseudo) Killing spinor, from which a timelike vector field is constructed, is of the form:
\[ ds^2 = -\left(dt +  {2 \sqrt{3} \over \chi} \mathcal{P} + e^{{\chi \over
    \sqrt{3}}t} \mathcal{Q}\right)^2 + e^{-{\chi \over \sqrt{3}}t} ds_\mathcal{M}^2 \ , \qquad A=\frac{dt}{2\sqrt{3}} +  {\mathcal{P} \over \chi}  + \frac{\sqrt{3}}{2}e^{{\chi \over
    \sqrt{3}}t} \mathcal{Q} \ , \]
where all $t$ dependence is \textit{explicit}, and the method to construct explicit solutions is the following:
\begin{itemize}
\item[1)] Take the base space $\mathcal{M}$ to be a four dimensional HKT geometry with metric $ds^2_\mathcal{M}$ and torsion tensor $H$;
\item[2)] The 1-form $\mathcal{P}$ is given by $ \mathcal{P}=-\star_4 H$;
\item[3)] Choose a 1-form $\mathcal{P}$ obeying the constraints
\[ \big( d \mathcal{Q} -2 \mathcal{P} \wedge \mathcal{Q} \big)^+ =0 \ ,
\qquad d \star_4 \mathcal{Q} + {16 \over \sqrt{3} \chi^3} d \mathcal{P} \wedge d \mathcal{P} =0 \ , \]
where $^+$ denotes the self-dual projection on the base-manifold $\mathcal{M}$, with positive orientation fixed with respect to the volume form ${\hat{\bf{e}}}^1 \wedge {\hat{\bf{e}}}^{\bar{1}} \wedge {\hat{\bf{e}}}^2 \wedge {\hat{\bf{e}}}^{\bar{2}}$. Note that one can always solve the second constraint; the general solution is given by
 \[
 \mathcal{Q} = {16 \over \sqrt{3} \chi^3} \star_4 (\mathcal{P} \wedge d \mathcal{P}) + \star_4 d \Phi \ ,
 \]
where $\Phi$ is a 2-form on $\mathcal{M}$. On substituting this expression back into the first constraint equation, one finds an equation constraining $\Phi$, which must be solved.
\end{itemize}

The Ricci scalar of the solution is
\[
\begin{array}{l}
\displaystyle{\mathcal{R}=\frac{5}{3}\chi^2+\frac{e^{\frac{2\chi}{\sqrt{3}} t}}{3}\left[\frac{(d\mathcal{P})^2}{\chi^2}-\frac{\chi^2}{2}e^{\frac{\chi}{\sqrt{3}} t}\mathcal{Q}^2  +\frac{3}{4}e^{\frac{2\chi}{\sqrt{3}} t}(d\mathcal{Q}-2\mathcal{P}\wedge \mathcal{Q})^2\right]} \ , \end{array}
\]
where the norms are computed with respect to the $t$-independent 
base space metric. Therefore, the $t$-dependence of
the Ricci scalar can be read directly from the above expression.
Thus, in particular, for the solution to be regular at both
$t=\pm \infty$ we must require $\mathcal{Q}=0$ and $d\mathcal{P}=0$. In particular, this implies that the base space is conformally hyper-K\"ahler.

If one assumes that $\mathcal{M}$ is a conformally hyper-K\"ahler, then $d\mathcal{P}=0$. After some coordinate transformations, the solution can be cast in the form
\[
ds^2=-f^2(dt+\omega)^2+f^{-1}ds^2_{HK} \ , \ \ \ \
F=\frac{\sqrt{3}}{2}d \bigg( f(dt+\omega)  \bigg) \ ,\]
where 
\[ f^{-1}=H-\frac{\chi}{\sqrt{3}}t \ ,  \ \label{f}\]
and 
\[ \Delta_{HK} H=0 \ , \qquad
(d \omega)^+ =0 \ . \label{conditionscanonical}
\]
$dS_5$ is obtained by taking the hyper-K\"ahler space to be ${E}^4$, $H=const.$ and $\omega=0$. 

This form of the solution is exactly the form of the solutions of minimal ungauged supergravity with $G^+=0$, except for the linear term in $t$ which arises in $f^{-1}$ due to the cosmological constant. Thus we are led to the following theorem: 

\textit{Any solution of $D=5$ minimal de Sitter supergravity with a pseudo Killing spinor and a base space which is conformal to a hyper-K\"ahler manifold can be obtained from a ``seed'' solution of minimal ungauged supergravity simply by adding a linear time dependence to the harmonic function.} 

This result generalises an earlier result by Behrndt and Cvetic \cite{Behrndt:03}. Moreover it makes clear why we can superimpose certain solutions (like the BMPV black hole \cite{Klemm:01} or G\"odel type universes \cite{Behrndt:04}) with a positive cosmological constant. But it also suggests that solutions with $G^+=0$ do not generalise easily to de Sitter space. Most notably this includes the black ring. 

Hence, in contrast to the AdS theory, the de Sitter theory admits multi-black hole solutions (like the multi-BMPV de Sitter), in which one finds multiple black holes co-moving with the expansion of the universe. This latter solution can be considered as a five dimensional rotating generalisation of the Majumdar-Papapetrou de Sitter solution, found by Kastor and Traschen \cite{Kastor:93}.

As a second class of examples one can take $\mathcal{M}$ to be an HKT manifold admitting a tri-holomorphic Killing vector field $X$. This means that both the base space metric and the almost complex structures are preserved by Lie dragging along the integral lines of $X$. Such HKT manifolds have been classified \cite{Chave:96, Gauduchon:98} and their structure is completely specified in terms of a constrained 3-dimensional Einstein-Weyl geometry. Explicit solutions can then be obtained, the simplest of which takes the Einstein-Weyl geometry to be a round 3-sphere \cite{Grover:08}. But all these solutions are singular, as expected from the above analysis of the Ricci scalar. The outstanding question is if, for any of these solutions, these singularities have interesting interpretations in terms of either black hole or big bang/big crunch singularities. That remains to be seen.


\begin{theacknowledgments}
C.H. and W.S. would like to thank the hospitality of
DAMTP Cambridge. CFP is partially funded by FCT
through the POCI programme. The work of W.S. is supported in
part by the National Science Foundation under grant number PHY-0703017.
\end{theacknowledgments}



\bibliographystyle{aipproc}   

\bibliography{sample}

\IfFileExists{\jobname.bbl}{}
 {\typeout{}
  \typeout{******************************************}
  \typeout{** Please run "bibtex \jobname" to optain}
  \typeout{** the bibliography and then re-run LaTeX}
  \typeout{** twice to fix the references!}
  \typeout{******************************************}
  \typeout{}
 }

\end{document}